\documentclass{llncs}
\usepackage{graphics}
\usepackage{epsfig}
\usepackage{algorithm}
\usepackage{algorithmic}
\usepackage{url}
\def\RETURN{\bf return }

\newtheorem{Pro}{Property}

\begin{document}

\pagestyle{empty} \mainmatter

\title{Clustering-Based Materialized View Selection \\ in Data Warehouses}

\titlerunning{Clustering-Based Materialized View Selection in Data Warehouses}
\author{Kamel Aouiche \and Pierre-Emmanuel Jouve \and J\'{e}r\^{o}me Darmont}
\institute{ERIC Laboratory -- University of Lyon 2\\
5, av. Pierre Mend\`{e}s-France\\
F-69676 BRON Cedex -- FRANCE\\
\email{\{kaouiche, pjouve, jdarmont\}@eric.univ-lyon2.fr}}

\maketitle

\begin{abstract}
Materialized view selection is a non-trivial task. Hence, its complexity must
be reduced. A judicious choice of views must be cost-driven and influenced by
the workload experienced by the system. In this paper, we propose a framework
for materialized view selection that exploits a data mining technique
(clustering), in order to determine clusters of similar queries. We also
propose a view merging algorithm that builds a set of candidate views, as well
as a greedy process for selecting a set of views to materialize. This selection
is based on cost models that evaluate the cost of accessing data using views
and the cost of storing these views. To validate our strategy, we executed a
workload of decision-support queries on a test data warehouse, with and without
using our strategy. Our experimental results demonstrate its efficiency, even
when storage space is limited.
\end{abstract}

\section{Introduction}

Among the techniques adopted in relational implementations of data warehouses
to improve query performance, view materialization and indexing are presumably
the most effective ones~\cite{vie03riz}. Materialized views are physical
structures that improve data access time by precomputing intermediary results.
Then, user queries can be efficiently processed by using data stored within
views and do not need to access the original data.
Nevertheless, the use of materialized views requires additional storage space
and entails maintenance overhead when refreshing the data warehouse.

One of the most important issues in data warehouse physical design is to select
an appropriate set of materialized views, called a configuration of views,
which minimizes total query response time and the cost of maintaining the
selected views, given a limited storage space. To achieve this goal, views that
are closely related to the workload queries must be materialized.

The view selection problem has received significant attention in the
literature. Researches about it differ in several points: (1) the way  of
determining candidate views; (2) the frameworks used to capture relationships
between candidate views; (3) the use of mathematical cost models \emph{vs.}
calls to the query optimizer; (4) view selection in the relational or
multidimensional context; (5) multiple or simple query optimization; and (6)
theoretical or technical solutions.

The classical papers in materialized view selection introduce a lattice
framework that models and captures dependency (ancestor or descendent) among
aggregate views in a multidimensional
context~\cite{bar97mat,far99eff,har96imp,kot99dyn,uch99pro}. This lattice is
greedily browsed with the help of cost models to select the best views to
materialize. This problem has been firstly addressed in one data cube and then
extended to multiple cubes~\cite{shu00mat}. Another theoretical framework
called the AND-OR view graph may also be used to capture the relationships
between views~\cite{gup97sel,goe99des,gup05sel,tho02ach,val02view}. The
majority of these solutions are theoretical and are not truly scalable. In
opposition to these studies, we exploit a query clustering involving similarity
and dissimilarity measures defined on the workload queries, in order to capture
the relationships existing between the candidate views derived from this
workload. This approach is scalable thanks to the low complexity of our
clustering (log linear regarding the number of queries and linear regarding the
number of attributes).

A wavelet framework for adaptively representing multidimensional data cubes has
also been proposed~\cite{smi04wav}. This method decomposes data cubes into an
indexed hierarchy of wavelet view elements that correspond to partial and
residual aggregations of data cubes. An algorithm greedily selects a
non-expensive set of wavelet view elements that minimizes the average
processing cost of the queries defined on the data cubes. In the same spirit,
Kotidis~\textit{et al.} proposed the Dwarf structure, which compresses data
cubes~\cite{sis02dwa}. Dwarf identifies prefix and suffix redundancies within
cube cells and factors them out by coalescing their storage. Suppressing
redundancy improves the maintenance and interrogation costs of data cubes.
These approaches are very interesting, but they are mainly focused on computing
efficient data cubes by changing their physical design. In opposition, we aim
at optimizing performance in relational warehouses without modifying their
design.

Other approaches detect common sub-expressions within workload queries in the
relational context~\cite{bar03sel,gol01opt,vie03riz,the04con}. The problem of
view selection consists in finding common subexpressions corresponding to
intermediary results that are suitable to materialize. However, browsing is
very costly and these methods are not truly scalable with respect to the number
of queries.

Finally, the most recent approaches are workload-driven. They syntactically
analyze the workload to enumerate relevant candidate views~\cite{agr00aut}. By
calling the query optimizer, they greedily build a configuration of the most
pertinent views. A workload is indeed a good starting point to predict future
queries because these queries are probably within or syntactically close to a
previous query workload. In addition, extracting candidate views from the
workload ensures that future materialized views will probably be used when
processing queries.

Our approach is also workload-driven. Its originality lies in exploiting
knowledge about how views can be used to resolve a set of queries to cluster
these queries together. For this purpose, we define the notion of query
similarity and dissimilarity in order to capture closely related queries. These
queries are grouped in the same cluster and are used to build a set of
candidate views. Furthermore, these candidate views are merged to resolve
multiple queries. This merging process can be seen as iteratively building a
lattice of views. The merging process time can be expensive when the number of
candidate views is high. However, we apply merging over candidate views present
in each cluster instead of the whole set of candidate views as
in~\cite{agr00aut}. This reduces the complexity of the merging process, since
the number of candidate views per cluster is significantly lower.

The remainder of this paper is organized as follows. We first present in
Section~\ref{sec:strategy} our materialized view selection strategy. Then, we
show in Section~\ref{sec:merge} how we build a candidate view configuration
through our merging process. Next, we detail in Section~\ref{sec:cost_model}
the cost models used for building the final configuration of views to
materialize. To validate our approach, we also present some experiments in
Section~\ref{sec:exp}. We finally conclude and provide research perspectives in
Section~\ref{sec:conclusion}.

\section{Strategy for materialized view selection}\label{sec:strategy}

\begin{figure}[t]
{\centering \resizebox*{0.9\textwidth}{!}{\includegraphics{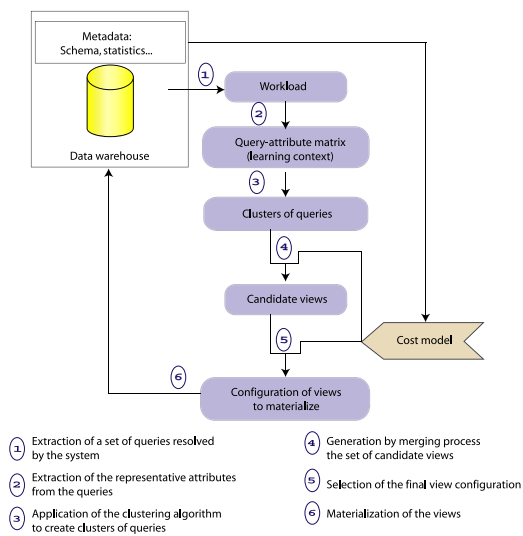}}
\par}
\caption{Strategy of materialized view selection}\label{fig:architecture}
\end{figure}

The architecture of our materialized view selection strategy is depicted in
Figure~\ref{fig:architecture}. We assume that we have a workload composed of
representative queries for which we want to select a configuration of
materialized views in order to reduce their execution time. The first step is
to build, from the workload, a context for clustering. This context is modelled
as a matrix having as many lines as the extracted queries and as many columns
as the extracted attributes from the whole set of queries. We define similarity
and dissimilarity measures that help clustering together relatively similar
queries. We apply a merging process  on each query cluster to build a
configuration of candidate views. Then, the final view configuration is created
with a greedy algorithm. This step exploits cost models that evaluate the cost
of accessing data using views and the cost of their storage.

\subsection{Query workload analysis}

The workloads we consider are sets of GPSJ (Generalized
Projection-Selection-Join) queries. A GPSJ query $q$ is composed of joins,
selection predicates and aggregations. As such, it may be expressed in
relational algebra over a star schema as follows: $q= \pi_{G,M}\sigma_{S}(F
\bowtie D_{1} \bowtie D_{2} \bowtie \dots \bowtie D_{d})$, where $S$ is a
conjunction of simple range predicates on dimension table attributes, $G$ is a
set of attributes from dimension tables $D_{i}$ (grouping set), and $M$ is a
set of aggregated measures each defined by applying aggregation operator to a
measure in fact table $F$. For example, query $q_{1}$ in
Figure~\ref{fig:charge} may be expressed as follows: $q_{1}=
\pi_{sales.time\_id,sum(quantity\_sold)}\sigma_{fiscal\_day=2}(sales \bowtie
times)$.

The first step consists in extracting from the workload the attributes that are
representative of each query. We mean by representative attributes those that
are present in \textbf{\texttt{Where}} (join and selection predicate
attributes) and \textbf{\texttt{Group by}} clauses. We also save for each query
their aggregation operators and joined tables. A query $q_{i}$ is then seen as
a line in a matrix composed of cells that correspond to the representative
attributes. The general term $q_{ij}$ of this matrix is set to 1 if the
extracted attribute is present in the query and to 0 otherwise. This matrix
represents our clustering context. Moreover, we store in an appendix matrix the
existing associations between the join attributes and queries, in the same
manner. We illustrate this step by an example: from the workload shown in
Figure~\ref{fig:charge}, we build the clustering context depicted in
Figure~\ref{tab:matrice_exemple}.

\begin{figure}[t]
\begin{center}
\begin{scriptsize}
\begin{tabular}{cp{11cm}}
\hline ($q_{1}$) & \textbf{select} sales.time\_id, sum(quantity\_sold)
\textbf{from} sales, times \\ & \textbf{where} sales.time\_id = times.time\_id
\textbf{and} times.fiscal\_day = 2
\\ & \textbf{group by} sales.time\_id;\\
($q_{2}$) & \textbf{select}  sales.prod\_id, sum(amount\_sold) \textbf{from}
sales, products, promotions \\ & \textbf{where} sales.prod\_id =
products.prod\_id \textbf{and} sales.promo\_id = promotions.promo\_id
\textbf{and} promotions.promo\_category = `newspaper'
\\ & \textbf{group by }sales.prod\_id;\\

($q_{3}$) & \textbf{select}  sales.cust\_id, sum(amount\_sold)
 \textbf{from} sales, customers, products, times \\
& \textbf{where} sales.cust\_id = customers.cust\_id \textbf{and}
sales.prod\_id = products.prod\_id \textbf{and} sales.time\_id = times.time\_id
\textbf{and} times.fiscal\_day = 3 \textbf{and} customers.cust\_marital\_status
=`single' \textbf{and} products.prod\_category =`Women' \\ & \textbf{group by}
sales.cust\_id;\\
 \dots &  \\ \hline
\end{tabular}
\end{scriptsize}
\caption{Example of workload} \label{fig:charge}
\end{center}

\begin{scriptsize}
\begin{tabular}{c|c|c|c|c|c|c|c|c|c|c|c|c| c l l}

\cline{2-13}

& $a_{1}$ & $a_{2}$ & $a_{3}$ & $a_{4}$ & $a_{5}$ & $a_{6}$ & $a_{7}$ & $a_{8}$  & $a_{9}$ & $a_{10}$ & $a_{11}$ & $a_{12}$ &  & $a_{1}$ times.time\_id & $a_{2}$ times.fiscal\_day \\

\cline{1-13}

\multicolumn{1}{|c|}{$q_{1}$} & 1 & 1 & 1 & 0 & 0 & 0 & 0 & 0 & 0 & 0 & 0 & 0 & & $a_{3}$ sales.time\_id & $a_{4}$ products.prod\_id \\

\cline{1-13}

\multicolumn{1}{|c|}{$q_{2}$} & 0 & 0 & 0 & 1 & 0 & 1 & 1 & 1 & 1 & 0 & 0 & 0 & & $a_{5}$ products.prod\_category & $a_{6}$ sales.promo\_id \\

\cline{1-13}

\multicolumn{1}{|c|}{$q_{3}$} & 0 & 0 & 0 & 1 & 0 & 1 & 1 & 1 & 1 & 1 & 1 & 1 & & $a_{7}$ promotions.promo\_id  & $a_{8}$ sales.prod\_id \\

\cline{1-13}

\multicolumn{1}{|c|}{$..$}&  &  &  &  &  &  &  &  &  & & & & & $a_{9}$
promotions.promo\_category  & $a_{10}$ sales.cust\_id \\ \cline{1-13}

\multicolumn{1}{|c|}{$..$}&  &  &  &  &  &  &  &  &  & & & & & $a_{11}$
customers.cust\_marital\_status  & $a_{12}$ customers.cust\_id \\ \cline{1-13}

\end{tabular}
\end{scriptsize}
\caption{Example of clustering context}\label{tab:matrice_exemple}
\end{figure}

\subsection{Building the candidate view set}

In practice, it is hard to search all the views that are syntactically relevant
(candidate views) from the workload queries, because the search space is very
large~\cite{agr00aut}. To reduce the size of this space, we propose to cluster
the queries. Indeed, we group in a same cluster all the queries that are
closely similar. Closely similar queries are queries having a close binary
representation in the query-attribute matrix. Two closely similar queries can
be resolved by using only one materialized view. Used within a clustering
process, the similarity and dissimilarity measures defined in the next section
ensures that queries within the same cluster  strongly relate to each other
whereas  queries from different clusters are significantly distant to each
other.

\subsubsection{Similarity measure.}

Let $QA$ be a query-attribute matrix that has a set of queries $Q=\{q_{i},
i=1..n\}$ as rows and a set of attributes $A=\{ a_{j}, j=1..p \}$ as columns.
The value $q_{ij}$ is equal to 1 if attribute $a_{j}$ is extracted from query
$q_{i}$. Otherwise, $q_{ij}$ is equal to 0. We describe query $q_{i}$ by a
vector of $p$ values $q_i=[ q_{i1}, ..., q_{ip}]$. These $p$ values describe
respectively the presence ($q_{ij}=1$) or absence ($q_{ij}=0$) of attribute
$a_{j}$. This description model helps comparing two queries. Then, for example,
we can consider queries $q_1$ and $q_2$ as closely similar if vectors
$[q_{11},..., q_{1p}]$ and $[q_{21},..., q_{2p}]$ have the majority of their
cells equal. This introduces the notion similarity and dissimilarity between
queries.


\subsubsection{Similarity and dissimilarity between queries.}

We define the notion of similarity and dissimilarity between queries  by two
functions $\delta_{sim_{j}}(q_{k},q_{l})$  and
$\delta_{dissim_{j}}(q_{k},q_{l})$ that measure the similarity between two
queries $q_{k}$ and $q_{l}$ with respect to attribute $a_j$.

$$
\delta_{sim_{j}}(q_{k},q_{l})=\left\{
\begin{tabular}{l}
$1$ if $q_{kj}=q_{lj}=1$\\
$0$ otherwise
  \end{tabular} \right.
$$

This first function defines the notion of similarity between $q_k$ and $q_l$
following attribute $a_j$: two queries $q_k$ and $q_l$ are considered similar
regarding attribute $a_j$ if and only if $q_{kj}=q_{lj}=1$, i.e., attribute
$a_j$ is extracted from both queries.

$$
\delta_{dissim_{j}}(q_{k},q_{l})=\left\{
\begin{tabular}{l}
$0$ if $q_{kj}=q_{lj}$\\
$1$ if $q_{kj} \neq q_{lj}$
 \end{tabular} \right.
$$

This second function defines the notion of dissimilarity between queries $q_k$
and $q_l$ according to attribute $a_{j}$:  two queries $q_k$ and $q_l$ are
considered dissimilar according to attribute $a_{j}$ if only and if $q_{kj}
\neq q_{lj}$, i.e., if one and only one of the queries does not contain
$a_{j}$. Note that there is not a complete symmetry between the notion of
similarity and dissimilarity:  non similar queries according to an attribute
are not necessarily dissimilar according to this attribute. For example, let
$q_1$ and $q_2$ be queries such that $q_{1j}=0$ and $q_{2j}=0$, respectively.
Then we have $\delta_{sim_{j}}(q_{1},q_{2})=0$ ($q_1$ and $q_2$ are not
considered similar) and $\delta_{dissim_{j}}(q_{1},q_{2})=0$ ($q_1$ and $q_2$
are not considered dissimilar). This absence of full symmetry underlines the
fact that the absence of the same attribute in two queries does not give an
element of similarity or dissimilarity between these queries.

These measures can be extended to an attribute set $A=\{ a_{1},\dots,a_{p}\}$
such that we get the degree of global similarity and dissimilarity between two
queries: $sim(q_{k},q_{l})=\sum_{j=1}^{p}\delta_{sim_{j}}(q_{k}, q_{l})$ and
$dissim(q_{k}, q_{l}) = \sum_{j=1}^{p} \delta_{dissim_{j}}(q_{k}, q_{l})$,
where $0\leq sim(q_{k},$ $ q_{l}) \leq p$ and $ 0\leq dissim(q_{k}, q_{l}) \leq
p$. Hence, the closer $sim(q_{a},q_{b})$ (resp. $dissim(q_{a},$  $q_{b})$) is
to $p$ the more $q_{a}$ and $q_{b}$ can be considered globally similar (resp.
dissimilar).

\subsubsection{Similarity and dissimilarity between query sets.}

As we do for two queries, we introduce two functions that take into account the
degree of similarity and dissimilarity between two query sets. A set of queries
(subset of $Q$) is denoted $C_{a}$. In order to translate the level of
similarity (resp. dissimilarity) between query sets, we use function
$Sim(C_{a},C_{b})$ (resp. $\quad$ $Dissim(C_{i},C_{j})$) that determines the
number of similarities (resp. dissimilarities) between two different sets of
queries $C_{a}$ and $C_{b}$ ($C_{a}\neq C_{b}$): $$Sim(C_{a},C_{b}) =
\sum_{q_{k}\in C_{a},q_{l}\in C_{b}}sim(q_{k},q_{l})$$ $$Dissim(C_{a},C_{b}) =
\sum_{q_{k}\in C_{a},q_{l}\in C_{b}}dissim(q_{k},q_{l})$$ where $0 \leq
Sim(C_{a},C_{b}) \leq card(C_{a}) \times card(C_{b}) \times p$ and $0 \leq
Dissim (C_{a},C_{b})  \leq card(C_{a}) \times card(C_{b}) \times p$. Hence, the
closer $Sim(C_{a},C_{b})$ (resp. $Dissim(C_{a},C_{b})$) is to $card(C_{a})
\times card(C_{b}) \times p$ the more $C_{a}$ and $C_{b}$ can be considered
similar (resp. dissimilar).

\subsubsection{Similarity and dissimilarity within a query set.}

The notion of similarity (resp. dissimilarity) within a query set corresponds
to the number of similarities (resp. dissimilarities) between queries of a same
set $C_{a}$. It consists of an extension of the similarity and dissimilarity
functions defined between query sets: $Sim(C_{a})= \sum_{q_{k}\in
C_{a},q_{l}\in C_{a}, k<l}sim(q_{k},q_{l}) \textrm{ and } Dissim(C_{a}) =
\sum_{q_{k}\in C_{a},q_{l}\in C_{a},k<l}dissim(q_{k},q_{l})$, where $0\leq
Sim(C_{a})\leq \frac{card(C_{a}) \times (card(C_{a})-1) \times p}{2}$ and
$0\leq Dissim(C_{a})\leq \frac{card(C_{a}) \times (card(C_{a})-1) \times
p}{2}$. Hence, the close $Sim(C_{a})$ (resp. $Dissim(C_{a})$) is to
$\frac{card(C_{a}) \times (card(C_{a})-1) \times p}{2}$ the more $C_{a}$
contains queries that are globally similar (resp. dissimilar).

\subsubsection{Query clustering.}

Clustering involves the determination of groups of objects (here: queries) that
reveal the the internal structure of data. These groups must be such as they
are composed of objects with high similarity and objects from different
clusters present a high dissimilarity.

Let us consider clustering $P_{h}$ of a query set, a quality measure of this
clustering can be built as follows:

\begin{scriptsize}
\begin{eqnarray*}
    Q(P_{h}) & = & \sum_{\begin{tabular}{c}
      a=1..z, \\
      b=1..z,a $<$ b \\
    \end{tabular}}Sim(C_{a},C_{b})+
    \sum_{a=1}^{z}Dissim(C_{a}) \label{eq:ker}
\end{eqnarray*}

$0  \leq Q(P_{h})  \leq \sum_{i=1..z,j=1..z,i<j} card(C_{i}) \times card(C_{j})
\times p + \sum_{i=1}^{z}\frac{card(C_{i}) \times (card(C_{i})-1) \times p}{2}$
\end{scriptsize}$ $\\

This measure permits to capture the natural aspect of a clustering. Hence,
$Q(P_{h})$ measures simultaneously similarities between queries within the same
cluster and dissimilarities between queries within different clusters. Thus,
$Q(P_{h})$ evaluates simultaneously the homogeneity of clusters as well as the
heterogeneity between clusters. Therefore, the clustering presenting a high
intra-cluster homogeneity and a high inter-cluster disparity has a weak value
of $Q(P_{h})$ and thereby appears as the most natural.


Jouve and Nicoloyannis proposed such a solution in the Kerouac
clustering algorithm and its associated clustering quality
measure~\cite{jou03ker}. We have chosen this algorithm because it has several
interesting  properties: (1) its computational complexity is relatively low (log
linear regarding the number of queries and linear regarding the number of
attributes) ; (2) it can deal with a high number of objects (queries)  ; (3) it
can deal with distributed data~\cite{jou03new}.

\section{View merging process}\label{sec:merge}

If we materialize all the different views derived from the query
clusters obtained in the previous step, we can obtain a high number of
materialized views, especially if the number of queries within the workload is
high. A view configuration obtained this way would not be very relevant if the
storage space allotted by the data warehouse administrator was limited. Instead
of materializing each view, it is better to only materialize views that can
be used to resolve multiple queries.  To solve this problem,
we must enumerate the space of views that can be merged, determine how to
guide the merging process, and finally build the set of merged views.
View merging is relevant if the queries are strongly similar. As we cluster
together closely similar queries, it is logical to apply the merging process
on the set of queries present in each cluster. This significantly reduces the
number of possible combinations when merging views. We detail in the following
sections how we merge two views and then generalize this process to
many views.\\

\noindent \textbf{Merging of view couples.} The merging of two views must
ensure these conditions: (1) all queries resolved by each view must also be
resolved by the merged view, and (2) the cost of using the couple of views must
not be significatively greater than the cost obtained when using the merged
view.
Let $v_{1}$ and $v_{2}$ be a couple of views of the same cluster and $s_{11},
\dots, s_{1m}$ the selection predicates that are in $v_{1}$ and not in $v_{2}$.
In a dual way, let $s_{21}, \dots, s_{1n}$ be the selection conditions present
in $v_{2}$ and not in $v_{1}$. Merged view $v_{12}$ is obtained by applying
Algorithm 1.\\

\hbox{ \parbox{5.8cm} {\begin{scriptsize} \textbf{Algorithm 1}
\textit{Merge\_View\_Pair$(v_{1},v_{2}$)}

\begin{algorithmic}[1]

\STATE put $v_{1}$ and $v_{2}$ aggregation operations in $v_{12}$ operation
aggregations

\STATE put the union of projection and group by attributes $v_{1}$ and $v_{2}$
in projection and group by clause of $v_{12}$

\STATE put all attributes $s_{11}, \dots, s_{1m}$ and  $s_{21}, \dots, s_{1n}$
in the group by clause of $v_{12}$

\STATE put the selection predicates shared between $v_{1}$ and $v_{2}$ in the
selection predicate clause of $v_{12}$

\end{algorithmic}
\end{scriptsize}}

\hspace{0.4cm} \parbox{5.8cm} {\begin{scriptsize} \textbf{Algorithm 2
}\textit{Mergin\_View\_Generation}

\begin{algorithmic}[1]

\STATE  $M = V_{1}$

\FOR{($k=2$; $V_{k-1} \neq \emptyset$; $k++$)}

\STATE $C_{k} =$ \textbf{\texttt{View\_Gen}}($V_{k-1}$)

\STATE  $M  \leftarrow M \cup C_{k}$

\FORALL{(view $v \in M $)}

\STATE Remove the parents of $v$ from $M$

\ENDFOR

\ENDFOR

\STATE \RETURN $M$\\ $ $\\ $ $\\
\end{algorithmic}
\end{scriptsize}}}
$ $\\

The merging  of two views $v_{1}$ and $v_{2}$ is effective if $cost (v_{12})
\geq ((cost (v_{1}) + cost (v_{2}))*x)$. Cost computation is detailed in
Section~\ref{sec:cost_model}. The value of $x$ is fixed empirically by the
administrator. If it is small (resp. high), we privilege (resp. disadvantage)
view merging.


\begin{Pro}
The view obtained by merging views $v_{1}$ and $v_{2}$ is the smallest view
that resolves the query resolved by both $v_{1}$ and $v_{2}$.
\end{Pro}
\begin{small}

\begin{proof}
To show that the view obtained by merging views $v_{1}$ and $v_{2}$ is the
smallest view, we have to show that there is no view $v'_{12}$ such as the data
within $v'_{12}$ are also included within $v_{12}$. We denote respectively
views $v_{1}$, $v_{2}$ and $v_{12}$ $\pi_{G_{1},M_{1}}\sigma_{S_{1}}(F \bowtie
\dots)$, $\pi_{G_{2},M_{2}}\sigma_{S_{2}}(F \bowtie \dots)$ and
$\pi_{G_{12},M_{12}}\sigma_{S_{12}}(F \bowtie \dots)$,
respectively, where:\\
-- $G_{1}$, $G_{2}$ are respectively the attribute set of the group
by clause of views $v_{1}$ and $v_{2}$;\\
-- $S_{1}$, $S_{2}$ are respectively the attribute set of the
selection predicates of $v_{1}$ and $v_{2}$;\\
-- $G_{12} = G_{1} \cup G_{2} \cup (S_{1} \cup S_{2} - S_{1} \cap S_{2})$ is
the attribute set of the group by clause of merged view
$v_{12}$;\\
-- $S_{12} = S_{1} \cap S_{2}$ is the set of attribute selection predicates
within merged view $v_{12}$.

Note that sets $G_{12}$ and $S_{12}$ are obtained by applying lines 1 and 2 of
Algorithm~1. Let us now assume that the data in view $v'_{12}$, denoted
$\pi_{G'_{12},M'_{12}}\sigma_{S'_{12}}(F \bowtie \dots)$ are all in $v_{12}$.
This means that both of the following conditions hold: (1) $G_{12} \subset
G'_{12}$, (2) $S_{12} \supset S'_{12}$.

From the first condition, there is at least one attribute $x$ such that $x
\in G'_{12}$ and $x \notin G_{12}$. As we have $x \notin G_{12}$, then $x
\notin G_{1}$, $x \notin G_{2}$ and $x \notin S_{1} \cup (S_{2} - S_{1} \cap
S_{2})$ because $G_{12} = G_{1} \cup G_{2} \cup (S_{1} \cup S_{2} - S_{1} \cap
S_{2})$. As $x \notin G_{1}$ and $x \notin G_{2}$, then $x$ is not in any
clause of $v_{2}$. This means that $x \notin G'_{12}$, which contradicts
condition $x \in G'_{12}$.

From the second condition, there is at least one attribute $y$ such that $y \in
S_{12}$ and $y \notin S'_{12}$. As we have $y \in S_{12}$, then $y \in S_{1}$
and $y \in S_{2}$ because $S_{12} = S_{1} \cup S_{2}$. As $y \in S_{1}$ and $y
\in S_{2}$, then $y$ must be in all the predicates of the views obtained by
merging $v_{1}$ and $v_{2}$. This means that $y \in S'_{12}$, which contradicts
condition $y \notin S'_{12}$.
\end{proof}
\end{small}

\noindent \textbf{Merged view generation algorithm.} The algorithm of view
generation by merging is similar to algorithms searching for frequent itemsets.
A frequent itemset lattice looks like a lattice of  views within a given cluster.
The lattice nodes represent the space of views obtained by merging. \\

\hbox{ \parbox{5.8cm}{\begin{scriptsize} \textbf{Algorithm 3 }\textit{Function
View\_Gen($V_{k-1})$}

\begin{algorithmic}[1]

\STATE  $C_{k}=\emptyset$

\FORALL{(view $v \in V_{k-1}$)}

\FORALL{(view $u \in V_{k-1}$)}

\IF {($v[1]=u[1] \wedge \dots \wedge v[k-2]=u[k-2] \wedge v[k-1] < u[k-1]$)}

\STATE $c =$ \textbf{\texttt{Merge\_View\_Pair}} ($v$,$u$)

\IF {($cost (c) \geq ((cost (v) + cost (u))*x)$)}

\STATE $C_{k} = C_{k} \cup \{c\}$  \

\ENDIF

\ENDIF

\ENDFOR

\ENDFOR

\STATE \RETURN $C_{k}$
\end{algorithmic}
\end{scriptsize} } \hspace{0.4cm}\parbox{5.8cm}{

\begin{scriptsize}
\textbf{Algorithm 4} \\ \textit{View\_Configuration\_Construction}

\begin{algorithmic}[1]

\STATE  $S \leftarrow \emptyset$ \REPEAT

\STATE $v_{max} \leftarrow \emptyset$

\STATE $F_{max} \leftarrow 0$

\FORALL{$v_{j} \in V-S$}

\IF {$F_{/S}(v_{j}) > F_{max}$}

\STATE $F_{max} \leftarrow F_{/S}(v_{j})$

\STATE $v_{max} \leftarrow v_{j}$

\ENDIF

\ENDFOR

\IF{$F_{/S}(v_{max}) > 0$}

\STATE $S \leftarrow  S \cup \{v_{max}\}$

\ENDIF

\UNTIL{($F_{/S}(v_{max}) \leq 0$ or $V-S=\emptyset$)}
\end{algorithmic}
\end{scriptsize}
}} $ $\\

The algorithm of view generation by merging (Algorithm~3) uses an iterative
ap\-proach by level to generate a new view. It  explores the view
lattice in breadth first. The input of the algorithm is $V_{1}$, a set of candidate views
extracted from a given cluster. This algorithm outputs a set of candidate views
obtained by merging. In the $k^{th}$ iteration, view set $V_{k-1}$ obtained by
merging the $k-1^{th}$ level's views from the lattice (computed in the last step) is
used to generate the set $C_{k}$ of $k$-candidate views. This set is added to set
$M$ (line~4). The parents of each view obtained by merging are then removed
from set $M$ (lines 5 to 7).

The function for view generation by merging
\textbf{\texttt{View\_Gen($V_{k-1}$)}}, called on line~3, takes as argument
$V_{k-1}$ and returns $C_{k}$. Two views $v$ and $u$ within $V_{k-1}$ form a
$k$-view $c$ if and only if they have ($k-2$) views in common. This is
expressed using a lexicographic order in the condition of line~3. We denote by
$v[1] \dots v[k-2] v[k-1]$ the merged views in the $k^{th}$ iteration that are
used to derive $v$. Function \textbf{\texttt{Merge\_View\_Pair}}($v$,$u$)
(Algorithm~1) called on line~5 of \texttt{View\_Gen} generates a new view $c$.
The condition of line~6 ensures, after generating a $k$-view from two
$k-1$-views, that the candidate view does not have a cost greater than the cost
of its parents.

\section{Cost models}\label{sec:cost_model}

The number of candidate views is generally as high as the input workload is
large. Thus, it is not feasible to materialize all the proposed views because
of storage space constraints. To circumvent these limitations, we use cost
models allowing to conserve only the most pertinent views. In most data
warehouse cost models~\cite{gol98met}, the cost of a query $q$ is assumed to be
proportional to the number of tuples in the view on which $q$ is executed. In
the following section, we detail the cost model that estimates the size of a
given view.

Let $ms(F)$ be the maximum size of fact table $F$, $|F|$ be the number of
tuples in $F$, $D_{i}\_ID$ be a primary key of dimension $D_{i}$, $|D_{i}\_ID|$
be the cardinality of the attribute(s) that form the primary key, and $N$ be
the number of dimension tables. Then, $ms(F)=\prod_{i=1}^{N}|D_{i}\_ID|$.

Let $ms(V)$ be the maximum size of a given view $v$ that has attributes
$a_{1},a_{2},\dots,a_{k}$ in its group by clause, where $k$ is the number of
attributes in $v$ and $|a_{i}|$ is the cardinality of attribute $a_{i}$. Then,
$ms(v)=\prod_{i=1}^{k}|a_{i}|$.

Golfarelli \textit{et al.}~\cite{gol98met} proposed to estimate the number of
tuples in a given view $v$ by using Yao's formula~\cite{yao77app} as follows:\\
$|v|=ms(v) \times \left[ 1-\prod_{i=1}^{|F|}\frac{ms(F) \times
d-i+1}{ms(F)-i+1}\right], \textrm{where }  d=1- \frac{1}{ms(v)}.$ If
$\frac{ms(F)}{ms(v)}$ is sufficiently large,
then Cardenas' formula~\cite{car75ana} approximation gives:\\
$|v|=ms(v) \times \left( 1-\left(1-\frac{1}{ms(v)}\right)^{|F|}\right),
\textrm{where } d=1- \frac{1}{ms(v)}.$

Cardenas' and Yao's formulaes are based on the assumption that data is
uniformly distributed. Any skew in the data tends to reduce the number of
tuples in the aggregate view. Hence, the uniform assumption tends to
overestimate the size of the views and give a crude estimation. However, they
have the advantage to be simple to implement and fast to compute. In addition,
because of the modularity of our approach, it is easy to replace the cost model
module by another more accurate one.

From the number of tuples in $v$, we estimate its size, in bytes, as follows:
$size(v)= |v| \times \sum_{i=1}^{c}size(c_{i})$, where $size(c_{i})$ denotes the
size, in bytes, of column $c_{i}$ of $v$, and $c$ is the number of columns in
$v$.

\section{Objective functions}

In this section, we describe three objective functions to evaluate the
variation of query execution cost, in number of tuples to read, induced by
adding a new view. The query execution cost is assimilated to the number of
tuples in the fact table when no view is used or to the number of tuples in
view(s) otherwise. The workload execution cost is obtained by adding all
execution costs for each query within this workload.

The first objective function advantages the views providing more profit while
executing queries, the second one advantages the views providing more benefit
and occupying the smallest storage space, and the third one combines the first two in
order to select at first all the views providing more profit and then keep only
those occupying the smallest storage space when this resource becomes critical. The
first function is useful when storage space is not limited, the second one is
useful when storage space is small and the third one is interesting when
storage space is larger.

\subsection{Profit objective function}

Let $V=\{v_{1},...,v_{m}\}$ be a candidate view set, $Q=\{q_{1},...,q_{n}\}$ a
query set (a workload) and $S$ a final view set to build. The profit objective
function, noted $P$, is defined as follows:\\ $P_{/S}(v_{j})=\left(
C_{/S}(Q)-C_{/S \cup \{v_{j}\}}(Q)- \beta \:
C_{maintenance}({\{v_{j}\}})\right)$,
where ${v}_{j} \notin S$.\\

\begin{itemize}
\item $C_{/S}(Q)$ denotes the query execution cost when all views in $S$ are
used. If this set is empty, $C_{/\emptyset}(Q) = |Q| \times |F|$ because all
the queries are resolved by accessing fact table $|F|$. When a view $v_{j}$ is
added to $S$, $C_{/S \cup \{v_{j}\}}(Q)= \sum_{k=0}^{|Q|} C(q_{k},\{v_{j}\})$
denotes the query execution cost for the views that are in $S \cup \{v_{j}\}$.
If query $q_{k}$ exploits $v_{j}$, the cost $C(q_{k},\{v_{j}\})$ is then equal
to $C_{v_{j}}$ (number of tuples in $v_{j}$). Otherwise, $C(q_{k},\{v_{j}\})$
is equal to the minimum value between $|F|$ and values of $C(q_{k},\{v\})$
(executing cost of
$q_{k}$ exploiting $v \in S$ with $v \neq v_{j}$).\\

\item Coefficient $ \beta = |Q| \: p(v_{j})$ estimates the number of updates for
view $v_{j}$. The update probability $p(v_{j})$ is equal to $\frac{1}{number \:
of \: views}\frac{\%update}{\%query}$, where  $\frac{\%update}{\%query}$
represents the proportion of updating \emph{vs.} querying the data
warehouse.\\

\item $C_{maintenance}(\{v_{j}\})$ represents the maintenance cost for view
$v_{j}$.

\end{itemize}

\subsection{Profit/space ratio objective function}

If view selection is achieved under a space constraint, the profit/space
objective function $R_{/S}(v_{j})= \frac{P_{/S}(v_{j})}{size(v_{j})}$ is used.
This function computes the profit provided by $v_{j}$ in regard to the storage
space $size(v_{j})$ that it occupies.

\subsection{Hybrid objective function}

The constraint on the storage space may be relaxed if this space is relatively
large. The hybrid objective function $H$ does not penalize space--``greedy''
views if the ratio $\frac{remaining\_space}{storage\_space}$ is lower or equal
than a given threshold $\alpha$ ($ 0< \alpha \leq 1 $), where
$remaining\_space$ and $storage\_space$ are respectively the remaining space
after adding $v_{j}$ and the allotted space needed for storing all the views.
This function is computed by combining the two functions $P$ and $R$ as
follows:

$ H_{/S}(v_{j})= \left\lbrace
\begin{tabular}{ll}
$P_{/S}(v_{j})$ & if $\frac{remaining\_space}{storage\_space} > \alpha$, \\
$R_{/S}(v_{j})$ & otherwise. \\
\end{tabular}
\right.$

\section{View selection algorithm}

The view selection algorithm (Algorithm~4) is based on a greedy search within
the candidate view set $V$. Objective function $F$ must be one of the functions
$P$ or $R$  described previously. If $R$ is used, we add to the algorithm's
input the space storage $M$ allotted for views.

In the first algorithm iteration, the values of the objective function are
computed for each view within $V$.
The view $v_{max}$ that maximizes $F$, if it exists ($F_{/S}(v_{max})
> 0$), is then added to $S$. If $R$
is used, the whole space storage $M$ is decreased by the amount of space
occupied by $v_{max}$.

The function values of $F$ are then recomputed for each remaining view in
$V-S$ since they depend on the selected views present in $S$. This helps taking
into account the interactions that probably exist between the views.

We repeat these iterations until there is no improvement ($F_{/S}(v) \leq 0$)
or until all views have been selected ($V-S= \emptyset $). If function $R$ is
used, the algorithm also stops when storage space is full.

\section{Experiments}\label{sec:exp}

In order to validate our approach for materialized view selection, we have run
tests on a 1 GB data warehouse implemented within Oracle $9i$, on a Pentium 2.4
GHz PC with a 512 MB main memory and a 120 GB IDE disk. This data warehouse is
composed of the fact table \textbf{\texttt{Sales}} and five dimensions
\textbf{\texttt{Customers}}, \textbf{\texttt{Products}},
\textbf{\texttt{Times}}, \textbf{\texttt{Promotions}} and
\textbf{\texttt{Channels}}. We executed on our data warehouse a workload
composed of sixty-one decision-support queries involving aggregation operations
and several joins between the fact table and dimension tables. Due to space
constraints, the data warehouse schema and the detail of each workload query
are available at \url{http://eric.univ-lyon2.fr/~kaouiche/adbis.pdf}. Our
experiments are based on an ad-hoc benchmark because, as far as we know, there
is no standard benchmark for data warehouses. TPC-R~\cite{tpcr99} has no
multidimensional schema and does not qualify, for instance.

We first applied our selection strategy with the profit function. This function
gives us the maximal number of materialized views (twelve views) because it
does not specify any storage space constraint. This point gives us the upper
boundary of the storage space occupation. Then, we applied the profit/space
ratio and hybrid functions under a storage space constraint. We have measured
query execution time with respect to the percentage of storage space allotted for
materialized views. This percentage is computed from the upper boundary
computed when applying the profit function. This helps varying storage space
within a wider interval.


\begin{figure}[h]
\begin{minipage}[b]{.5\linewidth}
 \centering\epsfig{figure=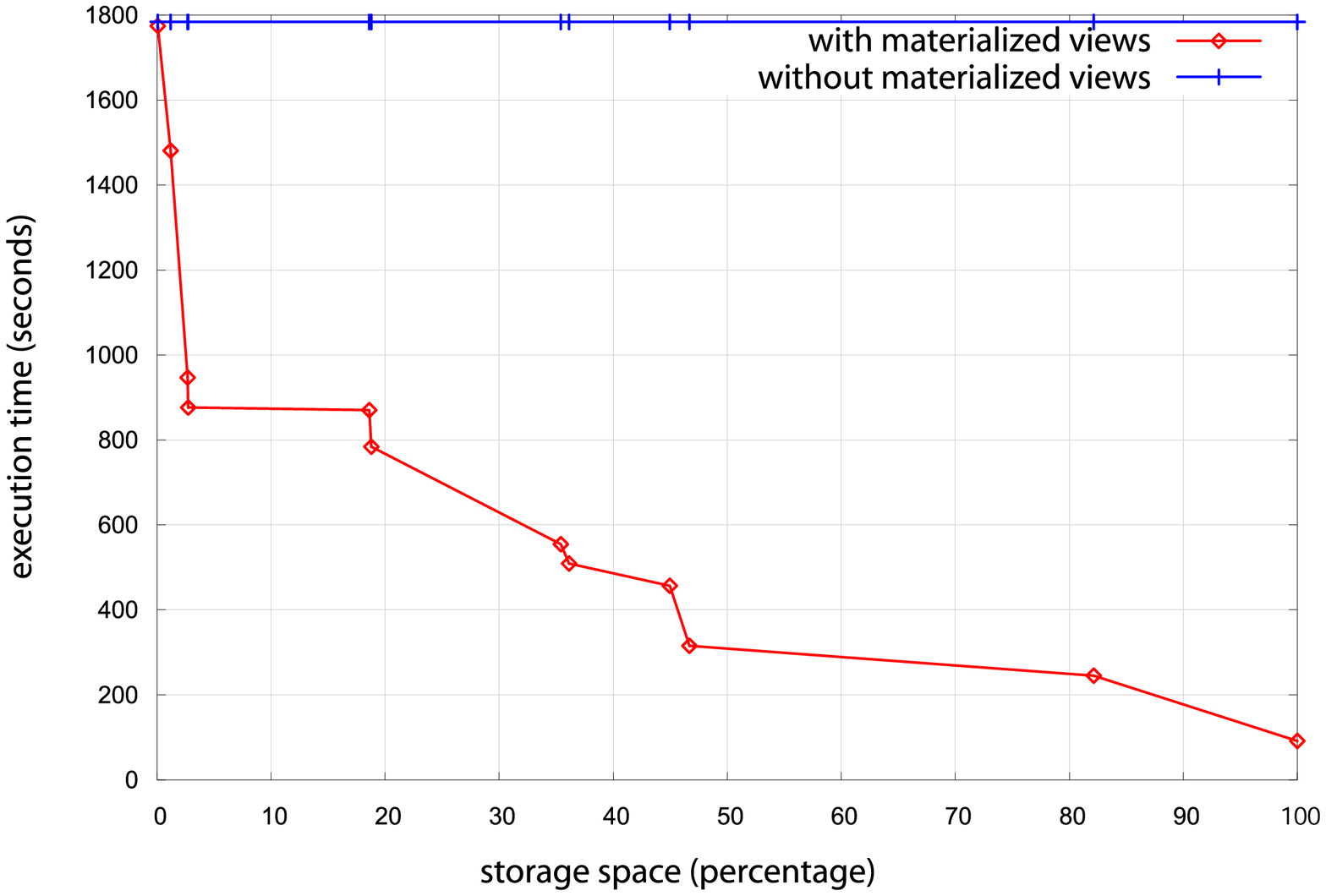,width=\linewidth}
\caption{Profit/space ratio function \label{fig:exp1}}
\end{minipage} \hfill
\begin{minipage}[b]{.5\linewidth}
 \centering\epsfig{figure=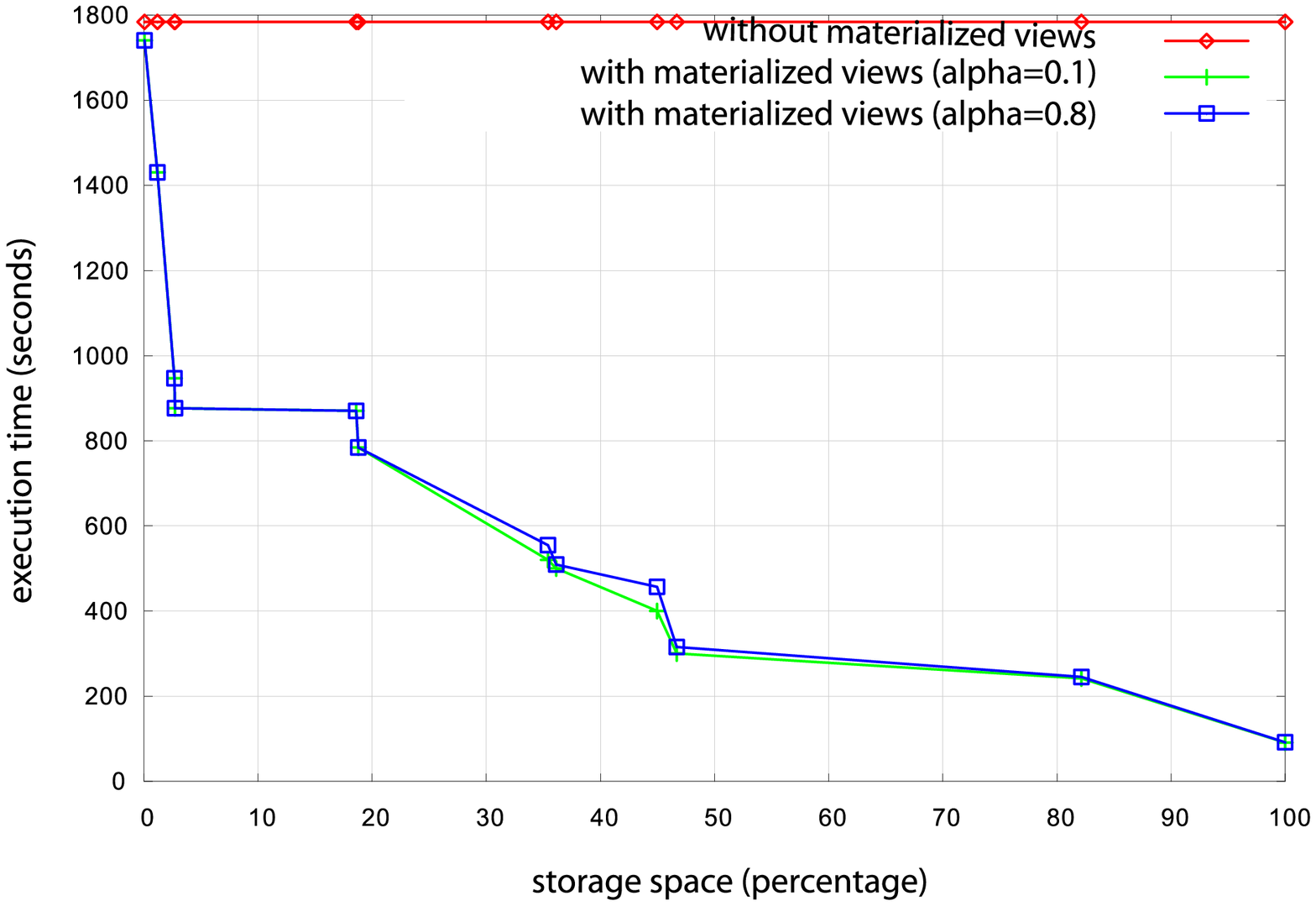,width=\linewidth}
 \caption{Hybrid function \label{fig:exp2}}
\end{minipage}
\end{figure}

\subsubsection{Ratio profit/space function experiment.}

We plotted in Figure~\ref{fig:exp1} the variation of workload execution time
with respect to the storage space allotted for materialized views. This figure
shows that the selected views improve query execution time. Moreover, execution
time decreases when storage space occupation increases. This is predictable
because we create more materialized views when storage space is large and
thereby better improve execution time. We also observe that the maximal gain is
equal to $94.86\%$. It is reached for a space occupation of $100\%$ (no
constraint on storage space). This case is also reached when using the profit
function, because it corresponds to the upper boundary.

\subsubsection{Hybrid function experiment.}

We repeated the previous experiment with the hybrid objective function. We
varied the value of parameter $\alpha$ between 0.1 and 1 by 0.1 steps. The
obtained results with $\alpha \in [0.1,0.7]$ and $\alpha \in [0.8,1]$ are
respectively equal to those obtained with $\alpha = 0.1$ and $\alpha = 0.8$.
Thus, we plotted in Figure~\ref{fig:exp2} only the results obtained with
$\alpha = 0.1$ and $\alpha= 0.7$. This figure shows that for percentage values
of space storage under 18.6\%, the hybrid function with $\alpha=0.1$ and
$\alpha=0.8$ behaves as the ratio function. When the storage space becomes
critical, the hybrid function behaves as the ratio profit/space function. On
the other hand, for the percentage values of storage space greater than 18.6\%,
the results obtained with $\alpha=0.8$ are slightly better than those obtained
with $\alpha=0.1$. This is explained by the fact that for the high values of
$\alpha$, the hybrid function chooses the views providing the most profit and
thereby improving the best the execution time. The maximal gain in execution
time observed for the values 0.1 and 0.8 of $\alpha$ is equal to $96\%$.

\begin{figure}[h]
{\centering \resizebox*{0.6\textwidth}{!}{\includegraphics{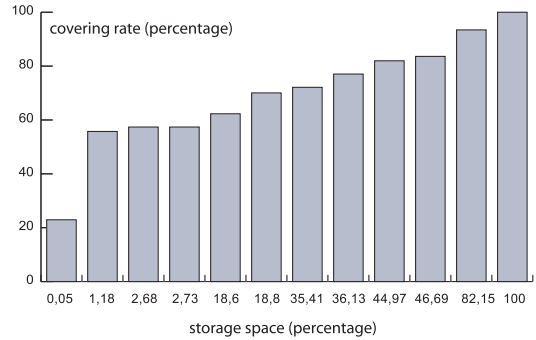}}
\par}
\caption{Query covering rate by the selected materialized views}
\label{fig:exp3}
\end{figure}


\subsubsection{Selected view pertinence experiment.}


In order to show if our strategy provides pertinent views for a given workload,
we measured the covering rate of the workload query results by the selected
views. We mean by covering rate the ratio between the number of queries
resolved from the materialized views and the total number of queries within the
workload. Thus, the highest the rate value, the most pertinent the selected
views. In this experiment, the percentage of storage space is also computed
from the upper boundary. We plotted in  Figure~\ref{fig:exp3} the covering rate
according to storage space occupation. This figure shows that the covering rate
increases with storage space. When storage space gets larger, we materialize
more views and thereby we recover more query results from these views. When
materializing all the views (100\% storage space occupation), all the data
corresponding to query results are recovered from the materialized views. This
shows that, without storage space constraint, the selected views are pertinent.
For example, for 0.05\% storage space occupation, 22.95\% of the query results
are recovered from the selected views. This shows that, even for a limited
storage space, our strategy helps building views that cover a maximum number of
queries. This experiment shows that materialized view selection based on
workload syntactical analysis is efficient to guarantee the exploitation of the
selected views by the workload queries.

\section{Conclusion}\label{sec:conclusion}

In this paper, we presented an automatic strategy for materialized view
selection in data warehouses. This strategy exploits the results of clustering
applied on a given workload to build a set of syntactically relevant candidate
views. Our experimental results show that our strategy guarantees a substantial
gain in performance. It also shows that the idea of using data mining
techniques for data warehouse auto-administration is a promising approach.

This work opens several future research axes. First, we are still currently
experimenting in order to better evaluate system overhead in terms of
materialized view building and maintenance. The maintenance cost is currently
derived from the query frequencies (Section~\ref{sec:cost_model}). We are
envisaging a more accurate cost model to estimate update costs. We also plan to
compare our approach to other materialized view selection methods. Furthermore,
it could be interesting to design methods that select both indexes and
materialized views, since these data structures are often used together. More
precisely, we are currently developing methods to efficiently share the
available storage space between indexes and views. Finally, our strategy is
applied on a workload that is extracted from the system during a given period
of time. We are thus performing static optimization. It would be interesting to
make our strategy dynamic and incremental, as proposed in~\cite{kot99dyn}.
Studies dealing with dynamic or incremental clustering may be exploited.
Entropy-based session detection could also be beneficial to determine the best
moment to run view reselection.

\bibliographystyle{abbrv}
\bibliography{mv}
\end{document}